\documentclass{ws-procs9x6}

\begin{document}

\title{The Transition between
Perturbative and Non-Perturbative QCD}


\author{\underline{A. Fantoni}}

\address{Laboratori Nazionali di Frascati dell'INFN, Via E. Fermi 40, \\ 
00044 Frascati (RM), Italy}

\author{S. Liuti}{

\address{University of Virginia, Charlottesville, Virginia 22901, USA}

\maketitle

\abstracts{
We study both polarized and unpolarized proton structure functions in the kinematical 
region of large Bjorken $x$ and four-momentum transfer of few GeV$^2$. 
In this region 
the phenomenon of parton-hadron duality takes place between the smooth continuation 
of the deep inelastic scattering curve and the average of the nucleon resonances. 
We present results on a perturbative-QCD analysis using all recent accurate data
with the aim of extracting the infrared behavior of the nucleon structure functions.}


\section{Introduction}
Parton-hadron duality is generally defined as 
the similarity between hadronic cross sections 
in the Deep Inelastic Scattering (DIS) region  and in the resonance region. 
It encompasses therefore a range of phenomena where one expects to observe a 
transmogrification from partonic to hadronic 
degrees of freedom, a question, the latter, at the very heart of Quantum ChromoDynamics 
(QCD).  

A number of experiments were conducted in the early days of QCD 
where the onset of parton-hadron duality was observed as the 
equivalence between the {\em continuation} of the smooth curve 
describing different observables from a wide variety of high energy and 
large momentum transfer 
reactions 
-- structure functions, sum rules, $R(s)$ for $e^+e^- \rightarrow$ hadrons, 
heavy meson decays... -- 
and the same observables in the low energy region 
characterized by low final state invariant mass values and 
resonance structure.
A fully satisfactory theoretical description of this phenomenon, that became to be 
accepted as a ``natural'' feature of hadronic interactions, is still nowadays 
very difficult to obtain. Recent progress both on the theoretical and experimental
side \cite{Jlab,Hermes}, has however both renovated and reinforced the hadronic 
physics community's interest in this subject as also demonstrated by the very lively
discussions among participants in the present Workshop.

In our contribution to the Workshop, we present evidence that standard
Perturbative QCD (PQCD) approaches in the large Bjorken $x$ region 
might not be adequate to describe
parton-hadron duality. In particular, by conducting an analysis of the most 
recent polarized and 
unpolarized inclusive electron scattering data, we unravel a discrepancy in the behavior
of the extracted power corrections from the DIS and resonance regions, respectively.
 
\section{Overview of data and QCD-based Interpretation} 
Parton-hadron duality being the idea that the outcome of any 
hard scattering process is determined by the 
initial scattering process among elementary constituents -- the
quarks and gluons -- independently from the hadronic phase  
of the reaction, is a well rooted concept in our current view
of all high energy phenomena. 
Bloom and Gilman (BG) duality \cite{BG} is the extension of this idea 
to a kinematical region characterized 
by lower center of mass energies of the hard scattering process.
Recently, more accurate experimental data have been collected that allow us to 
explore in detail this phenomenon. Besides the already mentioned 
new data on both inclusive electron-proton scattering \cite{Jlab}, and 
polarized inclusive electron-proton scattering \cite{Hermes}, several 
additional data sets and hadronic reactions were measured in the resonance region: 
polarized inclusive electron-proton scattering at Jefferson Lab kinematics 
\cite{Fat,Eg1},
$\tau \rightarrow \nu +$ hadrons \cite{SHI}, 
$\gamma p \rightarrow \pi^+ n$ \cite{Zhu,gao}, and, finally, 
inclusive electron-nucleus scattering \cite{Arr} (most of the recent data were
presented and discussed at this Workshop, see also \cite{Wally_rev} for 
a recent review of both theoretical and experimental results). 

A particularly interesting result was found 
in studies of inclusive reactions with no hadrons in the 
initial state, such as $e^+e^- \rightarrow$ hadrons, and 
hadronic $\tau$ decays \cite{SHI}. It was  
pointed out that, because of the truncation of the PQCD 
asymptotic series, terms including 
quark and gluon condensates play an increasing role as the center 
of mass energy of the process decreases.
Oscillations in the physical observables were then found
to appear if the condensates are calculated in an instanton 
background. Such oscillating structure, calculated in \cite{SHI} 
for values of the center of mass energy above the 
resonance region, is  
damped at high energy, hence warranting the onset of parton-hadron
duality. 

In what follows we examine a related question,
namely whether it is possible to extend the picture of duality 
explored in the higher $Q^2$ region \cite{SHI}, to the resonance region, 
or to the BG domain. 
A necessary condition is to determine 
whether the curve from the perturbative regime smoothly 
interpolates through the resonances, or whether, instead, violations of this
correspondence occur.
The latter would indicate that we are entering 
a semi-hard phase of QCD, where preconfinement effects might arise.
Our analysis of ``duality violations'' requires a sufficiently large and accurate set 
of data. We have applied it therefore to both the unpolarized and polarized 
inclusive measurements of proton structure functions in the resonance region.

\section{Monitoring the Transition between pQCD and npQCD}
We outline two important procedures for the study of parton-hadron
duality in structure functions: {\it i)} the {\em continuation} of DIS curve 
into the resonance 
region; {\it ii)} the {\em averaging} of the resonances.
Although these concepts are equivalently found in a number of different reactions, 
and in different channels (see {\it e.g.} \cite{Wally_rev,SHI}),
in this contribution we concentrate on
the proton structure functions $g_1$ and $F_2$, for polarized, and unpolarized 
electron scattering, respectively.
    
\subsection{Continuation of DIS Curve}
It is important to define exactly what one means by  ``continuation''
of the DIS curve, in order to be able to define whether parton-duality can be 
considered to be fulfilled.
The accuracy of current data   
allows us, in fact, to address the question of 
{\em what extrapolation from the large $Q^2$, or asymptotic
regime 
the cross sections in the resonance region should be compared to}.
In principle any extrapolation 
from high to low $Q^2$ is expected to be fraught with 
theoretical uncertainties ranging from the propagation of 
the uncertainty on $\displaystyle \alpha_S(M_Z^2)$
into the resonance region
to the appearance of different types
of both perturbative and power corrections in the low $Q^2$ regime. 
All of these aspects need therefore to be evaluated carefully.

Our approach applies to the {\em large Bjorken} $x$ behavior of inclusive
data. We therefore consider:

\vspace{0.2cm}
\noindent {\bf (a)} Non-Singlet (NS) Parton Distribution Functions (PDFs)
evolved at Next to Leading Order (NLO);
 
\noindent {\bf (b)} PQCD evolution using the scale $\approx Q^2(1-z)$ 
which properly takes into account integration over the parton's transverse momentum 
\cite{Brodsky,Roberts}; 

\noindent {\bf (c)}  Target Mass Corrections (TMCs).

\vspace{0.2cm}
\noindent 
We perform an extensive study of inclusive data in the resonance region by
extrapolating to this region all available parameterizations
of PDFs, which can be considered pure DIS, down to the 
measured ranges for $x$, $Q^2$, and final state invariant 
mass, $W^2=Q^2(1-x)/x + M^2$.
Our general approach for both the unpolarized structure function $F_2(x,Q^2)$, and the polarized 
one, $g_1(x,Q^2)$ is described in detail in \cite{BFL1}.  
An important point illustrated also in \cite{BFL1} is that 
the uncertainty due to the use of different
parameterizations can be taken into account by a band that is 
currently smaller than the experimental one in the region of interest.
A potential theoretical error in the
extrapolation of the ratios to low $Q^2$ could be generated
by the propagation of the error in $\alpha_S(M_Z^2)$. 
However, because 
at large $x$ perturbative evolution involves only Non-Singlet (NS) distributions, 
we expect it to affect minimally the extrapolation of 
the initial pQCD distribution, even to the low values of 
$W^2$ considered. 

The problem of resumming the
large logarithm terms arising at large $x$ was first noticed in a pioneering paper 
\cite{Brodsky}. There it was shown how this type of resummation can be taken care of 
by considering the correct definition of the upper limit of integration for the transverse 
momentum in the ladder diagrams defining the leading log approximation. This implies
replacing $Q^2$ with $\approx \widetilde{W}^2=Q^2(1-z)$, {\it i.e.} an invariant mass, 
in the evolution
equations. Such a procedure was used to obtain our results 
both in Refs.\cite{BFL1,SIMO1}, and in the current contribution.

Finally, TMCs are expected to be important at $W^2 \rightarrow M^2$.
Although TMCs are kinematic in nature and their 
contribution to the Operator Product Expansion (OPE) was calculated  
early on \cite{GP}, their effect on structure functions evolution 
cannot be evaluated 
straightforwardly at leading twist since a truncation of the twist
expansion brings inevitably to a mismatch between the Bjorken $x$ 
supports of the TM-corrected and the ``asymptotic'' results \cite{Mir}. 
The extent of this mismatch is, however, small  
in the DIS region thus rendering approximated treatments applicable \cite{GP}. 
In the resonance region the discrepancies arising in the \cite{GP} approach 
can be large. We adopt, therefore, the prescription of Ref.
\cite{Mir}, according to which 
the twist expansion for the standard moments Mellin 
of the structure function including TMCs is  
truncated consistently at the same order in $1/Q^2$
for both the kinematic terms and the dynamical higher twists. 
As noticed already in \cite{SIMO1}, this method applies 
for sufficiently small values of the expansion parameter 
$\approx 4M^2x^2/Q^2$. In addition, we have control on
the uncertainty which is a term of 
$O(1/Q^4)$.       

\subsection{Averaging Procedure}
Resonant data can be averaged over, according to different procedures 
We considered 
the following complementary methods:  
%
\label{average}
\begin{eqnarray}
& & I(Q^2)  =  \int^{x_{\mathrm{max}}}_{x_{\mathrm{min}}} 
 F_2^{\mathrm{res}}(x,Q^2) \; dx 
\label{Ires}
\\ \nonumber \\
& & M_n(Q^2)   =  {\displaystyle \int_0^1} dx \, \xi^{n-1} \, \frac{F_2^{\mathrm{res}}(x,Q^2)}{x} \, 
\, p_n 
\label{moment}
\\ \nonumber \\
& & F_2^{\mathrm{ave}}(x,Q^2(x,W^2))  =  F_2^{\mathrm{Jlab}}(\xi,W^2) 
\label{F2ave}
\end{eqnarray}
where $F_2^{\mathrm{res}}$ is evaluated using the experimental data  
in the resonance region
\footnote{Similar formulae hold for the polarized structure function, $g_1$.}.
In Eq.(\ref{Ires}), for each $Q^2$ value: 
$x_{\mathrm{min}}=Q^2/(Q^2+W_{\mathrm{max}}^2-M^2)$, and 
$x_{\mathrm{max}}=Q^2/(Q^2+W_{\mathrm{min}}^2-M^2)$. 
$W_{\mathrm{min}}$ and $W_{\mathrm{max}}$ delimit either 
the whole resonance region, {\it i.e.}
$W_{\mathrm{min}} \approx 1.1$ GeV$^2$, and $W_{\mathrm{max}}^2 \approx 4$ GeV$^2$, 
or smaller intervals within it.
In Eq.(\ref{moment}), $\displaystyle \xi$ 
is the Nachtmann 
variable \cite{NAC}, and $M_n(Q^2)$ are Nachtmann moments \cite{NAC}.
The r.h.s. of Eq.(\ref{F2ave}), $\displaystyle F_2^{\mathrm{Jlab}}(\xi,W^2)$, 
is a smooth fit to the resonant data \cite{Jlab}, valid
for $1< W^2 <4$ GeV$^2$; 
$F_2^{\mathrm{ave}}$ symbolizes the
average taken at the $\displaystyle Q^2 \equiv (x,W^2)$ of the data.

\section{Results}
After describing our program to address quantitatively all sources of 
theoretical errors started in \cite{SIMO1,BFL1}, in Fig.1 we 
present our main results on the extraction of 
the dynamical Higher Twist (HT) terms from the resonance region.  
A clear discrepancy marking perhaps a {\em breakdown of
the twist expansion} at low values of $W^2$ is seen for the 
unpolarized structure 
function, $F_2$ (upper panel). A comparison with other  results obtained 
in the DIS region \cite{MRST04} is also shown.  
For the polarized structure function, $g_1$, we added to our previous 
analysis data from \cite{Fat} at $Q^2 = 0.65, 1, 1.2$ GeV$^2$. 
In addition, we used the experimental values of the ratio 
$R=\sigma_L/\sigma_T$ from recent Jefferson Lab measurements in
the resonance region \cite{Kep_priv} which introduce an oscillation
around the original result of about $2\%$, well within the error bars. 
A complete presentation and discussion of these results along with
comparisons with other extractions \cite{Sidorov,Stam} will
be given in a forthcoming paper \cite{FL_inprep}.
From the figure one can see that although the trend seen in \cite{BFL1}
of a large violation of duality at $Q^2 \approx 1$ GeV$^2$ seems
to be confirmed,   
more polarized data at large $x$ are needed in order to 
draw definite conclusions.


\begin{figure}
\includegraphics[height=.5\textheight]{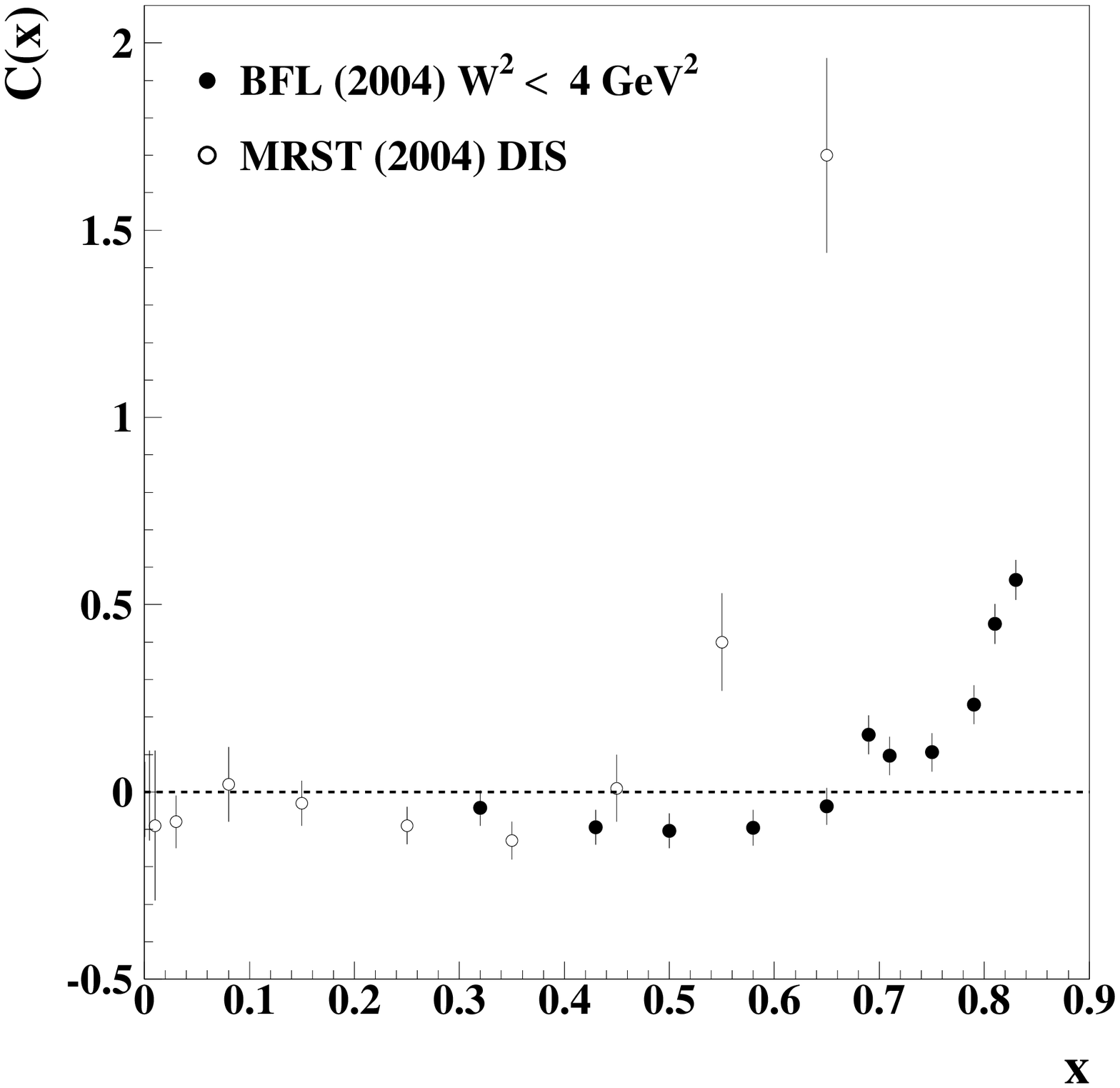}
  \includegraphics[height=.5\textheight]{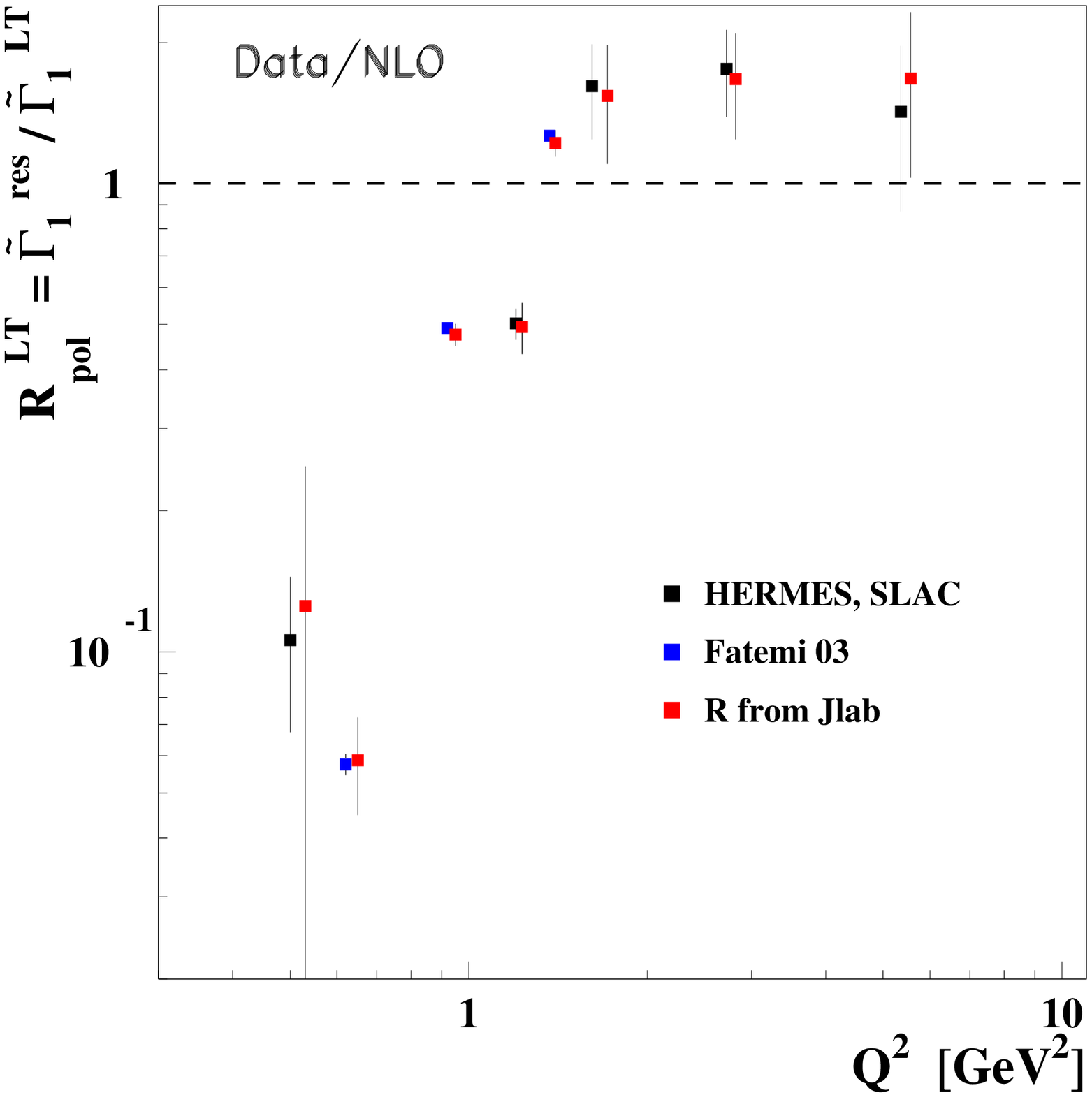}
  \caption{Upper panel: Comparison of HT contributions for the 
structure function $F_2$ 
in the DIS and resonance regions, respectively. 
The full circles are the values obtained 
in the resonance region \protect\cite{BFL1}.  
For $F_2$ these are compared with extractions using DIS data, 
from \protect\cite{MRST04}. 
Lower panel: ratio of the experimental data on $g_1$ 
\protect\cite{Hermes,Fat} and the PQCD extrapolation to
the resonance region \protect\cite{FL_inprep}.
Notice that the new results obtained using data 
from \protect\cite{Fat} agree 
with the trend of the Hermes data.}
\end{figure}


\vspace{0.3cm}
This work is partially  supported (S.L.) by the U.S. 
Department of Energy grant no. DE-FG02-01ER41200.


\begin{thebibliography}{9}

\bibitem{Jlab} I.~Niculescu {\it et al.},
  Phys.\ Rev.\ Lett.\  {\bf 85}, 1186 (2000). 

\bibitem{Hermes} A.~Airapetian {\it et al.}  [HERMES Collaboration],
  Phys.\ Rev.\ Lett.\  {\bf 90}, 092002 (2003)
    

\bibitem{BG} E.D. Bloom and F.J. Gilman, Phys. Rev. {\bf D4}, 2901 (1971); 
Phys. Rev. Lett. {\bf 25}, 1140 (1970). 

\bibitem{Fat} R. Fatemi {\it et al.}, Phys. Rev Lett. {\bf 91}, 222002 (2003). 

\bibitem{Eg1} Y. Prok, {\it these proceedings}. 

\bibitem{SHI}M.~A.~Shifman,  
arXiv:hep-ph/0009131; I.~I.~Bigi and N.~Uraltsev,
Int.\ J.\ Mod.\ Phys.\ A {\bf 16}, 5201 (2001); {\em these
proceedings}.

\bibitem{Zhu} L.Y. Zhu {\it et al.}, Phys. Rev. Lett. {\bf 91} 0220043 (2003);
Phys. Rev. {\bf C71} 044603 (2005).

\bibitem{gao} H. Gao, {\it these proceedings}. 

\bibitem{Arr} J. Arrington {\it et al.}, nucl-ex/0307012.

\bibitem{Wally_rev}
W.~Melnitchouk, R.~Ent and C.~Keppel,
 Phys.\ Rept.\  {\bf 406}, 127 (2005)

\bibitem{Brodsky}S.~J.~Brodsky and G.~P.~Lepage,
 SLAC-PUB-2447,
{\it Presented at Summer Inst. on Particle Physics, SLAC, Stanford, Calif., Jul 9-20, 1979}

\bibitem{Roberts} R.~G.~Roberts,
  Eur.\ Phys.\ J.\ C {\bf 10}, 697 (1999)

\bibitem{GP} H.~Georgi and H.~D.~Politzer,
  Phys.\ Rev.\ D {\bf 14}, 1829 (1976).

\bibitem{BFL1} N.~Bianchi, A.~Fantoni and S.~Liuti,
  Phys.\ Rev.\ D {\bf 69}, 014505 (2004)

\bibitem{SIMO1} S.~Liuti, R.~Ent, C.~E.~Keppel and I.~Niculescu,
  Phys.\ Rev.\ Lett.\  {\bf 89}, 162001 (2002)
  
\bibitem{Mir} J.~L.~Miramontes and J.~Sanchez Guillen,
  Z.\ Phys.\ C {\bf 41}, 247 (1988).

  \bibitem{NAC} O.~Nachtmann,
  Nucl.\ Phys.\ B {\bf 63}, 237 (1973).
  
\bibitem{MRST04} A.~D.~Martin, R.~G.~Roberts, W.~J.~Stirling and R.~S.~Thorne,
   Eur.\ Phys.\ J.\ C {\bf 35}, 325 (2004)

\bibitem{Sidorov} E.~Leader, A.~V.~Sidorov and D.~B.~Stamenov,
  Phys.\ Part.\ Nucl.\ Lett.\  {\bf 1}, 229 (2004), {\it and references therein}
       
\bibitem{Stam} D.~B.~Stamenov, {\it these proceedings}.

\bibitem{Kep_priv} C. Keppel, {\it private communication}.

\bibitem{FL_inprep} A. Fantoni and S. Liuti, {\it in preparation}.

\end{thebibliography}
\end{document}